\documentclass[prb,twocolumn,aps,showpacs,fixfloats]{revtex4}
\usepackage{graphicx}
\usepackage{bm}
\usepackage{amsmath,amssymb}
\usepackage{subfigure}
\usepackage{float}
\usepackage{latexsym}
\usepackage{epstopdf}
\usepackage{color}
\usepackage{enumerate}
\usepackage{pdfpages}

\newcommand{\s}{\scriptscriptstyle}

\begin{document}

\title {The shape of the Hanle curve in spin-transport structures in the presence of the ac drive  }

\author{R. C. Roundy$^{1}$, M. C. Prestgard$^{2}$, A. Tiwari$^{2}$, and M. E. Raikh$^{1}$ }

 \affiliation{$^{1}$Department of Physics and
Astronomy, University of Utah, Salt Lake City, UT 84112, USA \\
$^{2}$Department of Materials Science and Engineering, University of Utah, Salt Lake City, Utah 84112, USA
}

\begin{abstract}

Resistance between two ferromagnetic electrodes coupled to a normal channel
depends on their relative magnetizations. The spin-dependent component, $R$, of
the resistance changes with magnetic field, $B$, normal to the directions of
magnetizations. In the field of spin transport, this change, $R(B)$, originating from the Larmour spin precession, is called the Hanle curve. We demonstrate that the shape of the Hanle curve
evolves upon application
of an ac drive and
study this evolution theoretically as a function
of the amplitude, $B_1$, and frequency, $\omega$, of the drive.
If the distance between the electrodes, $L$, is smaller than the spin-diffusion length, $\lambda_s$, the prime effect of a weak circular-polarized drive is the shift of the center of the curve
to the value of $B$ for which the Larmour frequency, $\omega_L$,
is $\sim B_1^2/\omega$. Magnetic resonance at $\omega_L\sim \omega$ manifests itself in the derivative, $\frac{dR}{dB}$. For large $L \gg \lambda_s$ the ac drive affects the Hanle curve if the drive amplitude exceeds the spin relaxation rate, $\tau_s^{-1}$, i.e. at $B_1\tau_s \gtrsim 1$.
The prime effect of the drive is the elimination of a minimum in $R(B)$.
Linearly polarized drive has a fundamentally different effect on the Hanle curve, affecting not its shape, but rather its width.
\end{abstract}

\maketitle

\section{Introduction}
\begin{figure}
\includegraphics[width=77mm]{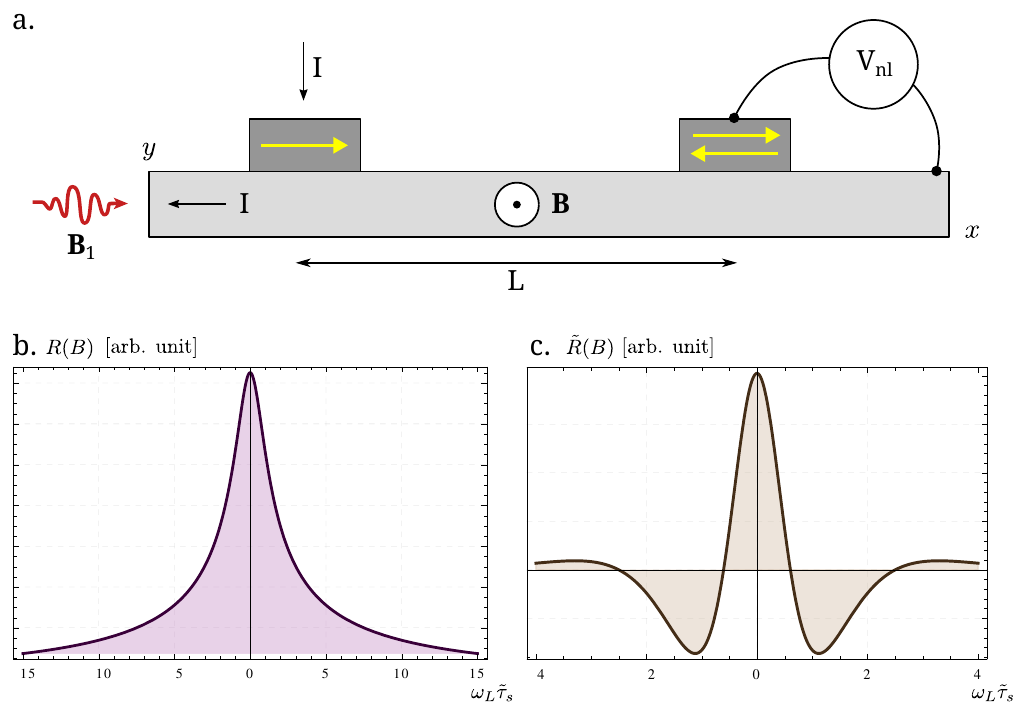}
\caption{[Color online] (a) Schematics of a standard spin-transport device.
Nonlocal resistance is defined as a voltage between the
channel and the right ferromagnet detector generated upon injecting the
current through the left ferromagnetic electrode.  While the polarized electrons
travel diffusively  the distance, $L$, their spin precesses in magnetic field, $B$, directed along the $z$--axis. The driving ac field, $B_1$ is either circularly polarized in the $x-y$--plane or linearly polarized along $x$. The shape of the Hanle curves
in (b)  short $L/\lambda_s =0.1$ and (c) long, $L/\lambda_s=2.4$, devices are calculated
from Eq. (\ref{propto}).  }
\label{device}
\end{figure}

In the past decade there has been  remarkable progress in the
fabrication of  lateral F-N-F  structures, see Fig. \ref{device},
which exhibit spin transport. In the pioneering experiment Ref. \onlinecite{vanWeesPioneering} the existence of spin transport in an Al strip was demonstrated by measuring a voltage, $V$,
generated between the strip and Co electrode upon injecting a current, $I$, through the other Co electrode.
The sign of voltage  could be reversed upon reversal of the relative magnetizations of the electrodes. Quantitative information about the spin transport was inferred from the dependence of the generated voltage on a weak external field, $B$, which caused the spin precession. In particular, it was observed that for the average spin precession angle $180^{\circ}$ the generated voltage changes the sign.

A theory for the $\frac{V}{I}=R(B)$ dependence,
i.e. for the Hanle profile,
was first developed
in  Refs.~\onlinecite{Silsbee1985}, \onlinecite{Silsbee1988}.
Following Ref. \onlinecite{vanWeesPioneering},
a concise derivation of the analytical result of  Refs.~\onlinecite{Silsbee1985}, \onlinecite{Silsbee1988}
goes as follows. Suppose that the magnetizations of the injector and detector are
directed along the $x$--axis, while the field, $B$, is directed along the $z$--axis. After a time, $t$, from the moment of injection the average $x$--projection of spin of a given electron is $S_x(t)=e^{-t/\tau_s}\cos\omega_Lt $, where $\omega_L=\gamma B$ is the Larmour frequency ($\gamma$ is the gyromagnetic ratio), and $\tau_s$ is the spin-flip time in the nonmagnetic material.
If the motion of electron between the electrodes is a 1D drift, then the times of arrival to
the detector are distributed as $P(t)=\frac{1}{\sqrt{4\pi Dt}}\exp\left[-(L-v_{\s d}t)^2/4Dt\right]$.
Here $L$ is the distance between the electrodes, see Fig. \ref{device}, $v_{\s d}$ is the drift velocity, and $D$ is the diffusion
coefficient. Then the nonlocal resistance, $R(B)$, is proportional $S_x(t)$ weighted with the distribution $P(t)$, i.e.
\begin{equation}
\label{R-dc}
R(B) = R_0 \int_0^\infty \frac{dt \cos(\omega_L t)}{\sqrt{4\pi Dt}}  \exp\left[-\frac{t}{\tau_s} -\frac{(L-v_{\s d}t)^2}{4Dt}\right].
\end{equation}
The prefactor $R_0$ is $B$-independent and is proportional to the product of polarizations of the injector
and detector.

The integral Eq. (\ref{R-dc}) contains four parameters of the device:
$D$, $\tau_s$, $L$, and $v_{\s d}$. In fact, the $B$ dependence of $R$ is governed by only two dimensionless combinations: $\omega_L{\tilde\tau}_s$, where ${\tilde \tau}_s$ is the renormalized spin-flip time
\begin{equation}
\tilde{\tau}_s=\frac{\tau_s}{1+\frac{v_0^2\tau_s}{4D}},
\end{equation}
and the dimensionless length
\begin{equation}
\tilde{L}=\frac{L}{\left(4D \tilde{\tau}_s\right)^{1/2}}.
\end{equation}
Besides, the integral can be evaluated analytically\cite{vanWeesPioneering,AnalyticalHanle}, and expressed in terms of the function $f(y)$ defined as
\begin{equation}
\label{f(y)}
f(y)=\int\limits_0^{\infty}\frac{ds}{s^{1/2}}\exp{\Bigl[-\frac{1}{s}-ys\Bigr]}=
\Bigl(\frac{\pi}{y}\Bigr)^{1/2}   \exp{\Bigl[-2y^{1/2}\Bigr]}.
\end{equation}
Then the $B$-dependence of the nonlocal  resistance is simply given by

\begin{multline}
\label{propto}
R(B)\propto f_{r}(y) =  {\text Re}f(y)  \\
=\Bigl(\frac{\pi}{\vert y \vert}\Bigr)^{1/2}\exp{\Bigl[-2\vert y\vert^{1/2}\cos\frac{\phi}{2}\Bigr]}
\cos\Bigl(\frac{\phi}{2}+2\vert y\vert^{1/2}\sin\frac{\phi}{2}\Bigr),
\end{multline}
where the absolute value, $|y|$, and the phase, $\phi$, of the complex argument, $y$, are defined as
\begin{equation}
\label{complex}
\vert y \vert =\tilde{L}^2\Bigl(1+\omega_L^2\tilde{\tau}_s^2\Bigr)^{1/2},~~~~\phi=\arctan \left(\omega_L\tilde{\tau}_s\right).
\end{equation}
It follows from Eq. (\ref{propto}) that there are two characteristic shapes of the Hanle curve, loosely speaking, short-device shape and long-device shape. They are illustrated in Fig.~\ref{device}.

With regard to experiments, Eq. (\ref{R-dc}) provides a remarkably accurate description of the Hanle curves measured in
various spin-transport devices. Both shapes of $R(B)$ have been  reported in many papers, see {\em e.g.} Refs.~
\onlinecite{AppelbaumSi,Appelbaum2008,Appelbaum,AnalyticalHanle,InvertedHanleSi,
Ge2013,Ge2014,
Crowell2005,Crowell2007,Wunderlich,GaAsKirczenow2013,AlGaAs2014,
graphene2007,graphene2009,WeesGraphene,graphene2013,graphene2014,
Crowell2010,Kuhlen2012,
GaN2012}.
Usually the value $\tau_s$ is inferred from $R(B)$, since $R(B)$ falls
off at values $\omega_L \sim \tau_s^{-1}$.
For example, in silicon-based\cite{AppelbaumSi,Appelbaum2008,Appelbaum,AnalyticalHanle,InvertedHanleSi}
and germanium-based\cite{Ge2013,Ge2014} structures the widths of the Hanle curves are $\sim 5$~mT, so that the values of $\tau_s$ is long, $\tau_s \sim  5$ ns.  Equally, in GaAs the Hanle curves are narrow\cite{Crowell2005,Crowell2007,Wunderlich,GaAsKirczenow2013,AlGaAs2014} with similar widths.
On the other hand, in
graphene-based  valves
\cite{graphene2007,graphene2009,WeesGraphene,graphene2013,graphene2014}
the Hanle curves are broad with widths  $\sim 100$~mT.
In InGaAs\cite{Crowell2010,Kuhlen2012} the widths are intermediate $\sim 20$~mT.

It should be noted that determination of {\em both} $\tau_s$ and the spin-diffusion length, $\lambda_s =(D\tau_s)^{1/2}$, from a single measured Hanle
curve is somewhat ambiguous,
in the sense, that the same $R(B)$ can be very well fitted with
two significantly different sets of $\tau_s$ and $\lambda_s$. To improve the accuracy of determination
of these parameters in Ref.~\onlinecite{Crowell2007} the Hanle curves for several values of $L$ were analyzed.

Overall, the excellent agreement of the experimentally measured Hanle profiles with theoretical prediction Eq.~(\ref{propto})
seems surprising, since the theory is based on a rather crude description of the spin dynamics of injected carriers. For example, this description completely neglects the
details of injection, such as geometry of electrodes. Modeling the transport as a purely
1D diffusion is also somewhat questionable\cite{Silsbee2007}.
On the other hand, a complete understanding of the domain of applicability and limitations of the drift-diffusion theory of spin transport seems crucial, since the
contemporary research on inverse spin Hall effect
\cite{Crowell2010,Wunderlich,LatestFert1,LatestFert2,HanleOrganicJapanese,Tiwari}
 and its possible applications in the logic devices contains the
drift-diffusion description at its core.
One way of testing the drift-diffusion theory, which have already been realized experimentally\cite{stripes,TwoBeams}, is to operate with spatially inhomogeneous spin-density profiles. For example in Ref. \onlinecite{TwoBeams} this profile was created using the interference of two laser beams.

In the present manuscript we suggest another ``knob" to test the drift-diffusion theory. Namely, we demonstrate that the Hanle profile can be manipulated by the ac drive. More specifically, we assume that,
in addition to a static field $B_0$,
an ac field in the $x$--$y$ plane is applied.  On  general grounds, one can expect that the ac drive suppresses the $R(B)$ response by affecting the steady precession $S_x(t)=\cos\omega_Lt$. It is also apparent that the  drive should make the most pronounced effect on $R(B)$ if the drive frequency, $\omega$, is comparable
to $\tilde \omega_L$ -- the  value corresponding to the width of the Hanle curve in the absence of drive.
For $\omega \gg \tilde \omega_L $
the ac field oscillates many times as an electron travels between the injector and detector, so that the effect of a weak drive with amplitude $\gamma B_1\ll \omega$ averages out. It is somewhat unexpected that, in
addition to a simple broadening, the drive gives rise to specific features in the shape of the Hanle curves.

Below we find and analyze the expression for $R(B)$ for
ac field with arbitrary amplitude and frequency for the case
when it is circularly polarized. The prime effect is the shift of
the center of $R(B)$ to the left or to the right depending whether the polarization of the drive is left or right.
We also analyze the evolution
of the Hanle curves with increasing drive for the case when the drive is linearly polarized.
In particular, we identify two peculiar regimes of the spin dynamics which are specific to linear polarization. They are realized when the drive is either very fast or very strong. We discuss how this dynamics
manifests itself in the Hanle profile.

\section{Dynamics of the Larmour spin precession in the presence of the ac drive}

To find the shape of the Hanle curve in the presence of the ac drive,
${\bm B}_1(t)$, it is necessary to solve the equation for the spin dynamics
\begin{equation}
\frac{d{\bm S}}{d t}+\gamma\Bigl({\bm B}+{\bm B}_1(t)\Bigr)\times {\bm S} =0
\end{equation}
with initial conditions $S_x(0)=1$, $S_y(0)=S_z(0)=0$.
Then the solution should be substituted into Eq. (\ref{R-dc}) instead of $\cos\omega_Lt$.
We assume that external field is directed along $z$, i.e.
${\bm B}= B_0{\bm k}$, while for the ac field lies in the $x$-$y$ -- plane. For this field we will consider the cases of circular and  linear polarization separately.
\subsection{Circular polarization}
It is important that the components of the ac drive
\begin{equation}
B_x=B_1
\cos(\omega t+\varphi)~~~~~B_y=B_1\sin(\omega t +\varphi)
\end{equation}
contain a random initial phase, $\varphi$. It emerges as a result of the randomness of the time moments at which electrons are injected from the electrode.
The nonlocal resistance should be averaged over this phase.

For circular polarization the dynamics of the spin components can be found exactly, since in the rotating frame
the ac field is static. We reproduce this textbook solution
to track the random phase, $\varphi$, which leads to averaging out   of certain contributions to $R(B)$.

In the rotating frame, $x'=x\cos\omega t+y\sin\omega t$,
 $y'=y\cos\omega t-x\sin\omega t$, the general solution of the
 Bloch equation has the Rabi form
\begin{multline}
\label{vectors}
{\bm S}'(t) = \left( {\bm S}_0 - \frac{({\bm H} \cdot {\bm S}_0) {\bm H}}{H^2} \right) \cos \gamma H t \\
+ \frac{{\bm H} \times {\bm S}_0}{H} \sin \gamma H t
+ \frac{({\bm H} \cdot {\bm S}_0) {\bm H}}{H^2},
\end{multline}
where the projections of the vector ${\bm H}$, which is the effective magnetic field in the rotating frame, are defined as $H_{x'} = B_1 \cos \varphi$, $H_{y'} = B_1 \sin \varphi$, and
$H_{z'} = B_0 - \frac{\omega}{\gamma}$. After implementing the initial condition $S_{x'}(0)=1$ it is instructive to
rewrite Eq. (\ref{vectors}) in components
\begin{widetext}
\begin{align}
\label{projections}
S_{x'}(t) &= \left(1 - \frac{B_1^2 \cos^2 \varphi}{B_1^2 + \left( B_0 - \frac{\omega}{\gamma}\right)^2} \right)
\cos \gamma H t +  \frac{B_1^2 \cos^2 \varphi}{B_1^2 + \left( B_0 - \frac{\omega}{\gamma}\right)^2}, \nonumber \\
S_{y'}(t) &= \frac{B_1^2 \cos \varphi \sin \varphi}{B_1^2 +
\left(B_0 - \frac{\omega}{\gamma} \right)^2} \left( 1- \cos \gamma H t\right) + \frac{B_0 - \frac{\omega}{\gamma}}{\sqrt{B_1^2 + \left(B_0 - \frac{\omega}{\gamma} \right)^2 }} \sin \gamma H t, \nonumber \\
S_{z'}(t) &= - \frac{B_1\left(B_0 - \frac{\omega}{\gamma} \right) \cos \varphi}{B_1^2 +
\left(B_0 - \frac{\omega}{\gamma} \right)^2} (1- \cos \gamma H t)
- \frac{B_1 \sin \varphi}{\sqrt{B_1^2 +
\left(B_0 - \frac{\omega}{\gamma} \right)^2}}
\sin \gamma H t.
\end{align}
\end{widetext}
It is now seen from Eq. (\ref{projections}) that $S_{z'}$ vanishes after averaging and so does the first term in $S_{y'}$. In the remaining terms the averaging amounts to the replacement of $\cos^2\varphi$ by $1/2$.

Going back to the laboratory system: $S_x=S_{x'}\cos\omega t
-S_{y'}\sin\omega t$, $S_y=S_{y'}\cos\omega t
+S_{x'}\sin\omega t$, we get
\begin{widetext}
\begin{align}
\label{Sx}
\langle S_{x} (t)\rangle &= \frac{B_1^2}{2 H^2} \cos \omega t
+  \frac{1}{2} \left( 1 - \frac{B_1^2}{2 H^2} + \frac{B_0-\frac{\omega}{\gamma}}{H} \right)\cos\left[ (\gamma H+\omega)t \right]
+
\frac{1}{2} \left( 1 - \frac{B_1^2}{2 H^2} - \frac{B_0-\frac{\omega}{\gamma}}{H} \right)\cos\left[ (\gamma H-\omega)t \right],  \\
\label{Sy}
\langle S_{y} (t)\rangle &= \frac{B_1^2}{2 H^2} \sin \omega t
+  \frac{1}{2} \left( 1 - \frac{B_1^2}{2 H^2} + \frac{B_0-\frac{\omega}{\gamma}}{H} \right)\sin\left[ (\gamma H+\omega)t \right]
-
\frac{1}{2} \left( 1 - \frac{B_1^2}{2 H^2} - \frac{B_0-\frac{\omega}{\gamma}}{H} \right)\sin\left[ (\gamma H-\omega)t \right].
\end{align}
\end{widetext}
We see the averaged dynamics of $S_x$ and $S_y$ represents oscillations
with driving frequency, $\omega$, which are modulated by the ``Rabi" envelope with frequency\cite{Rabi}
\begin{equation}
\gamma H=\sqrt{(\omega-\gamma B_0)^2 + (\gamma B_1)^2}.
\end{equation}
Thus, while the Rabi oscillations in $S_z(t)$ do not survive averaging
over initial phase, $\varphi$, they are still present in the averaged
dynamics of $S_x$ and $S_y$.
In the next section we study how this dynamics manifests itself in nonlocal resistance.


\section{Nonlocal resistance}

Three contributions to $S_x$ in Eq. (\ref{Sx}) give rise to three terms in the nonlocal resistance, $R(B)$. It is convenient to express $R(B)$ through the same function, $f_r(y)$, which describes the Hanle shape in the absence of drive and is defined by Eq. (\ref{f(y)}).
One finds
\begin{multline}
\label{three-terms}
\hspace*{-1cm}
R(B) \propto \Biggl\{\frac{B_1^2}{2 H^2} f_r\left(y_{\s \omega} \right)
 +  \frac{1}{2} \left( 1 - \frac{B_1^2}{2 H^2} + \frac{B_0-\frac{\omega}{\gamma}}{H} \right)f_r\left(y_{\s \omega + \gamma H} \right) \\
+
\frac{1}{2} \left( 1 - \frac{B_1^2}{2 H^2} - \frac{B_0-\frac{\omega}{\gamma}}{H} \right)f_r\left(y_{\s - \omega +\gamma H}\right)\Biggr\}.
\end{multline}
Here the arguments $y_{\s \omega}$ and  $y_{\s \pm \omega + \gamma H}$ are defined by Eq. (\ref{complex}) with $\omega_L$  replaced
by $\omega$ and $\pm \omega +\gamma H$, respectively.
It is convenient to analyze the shape of the Hanle curves for short and
long devices separately.

\subsection{Small distance between the electrodes}

In the limit of small $\tilde{L}\ll 1$ the function Eq. (\ref{propto}) for nonlocal resistance in the absence of drive simplifies to
\begin{equation}
\label{smallL}
R(B)\propto \frac{\sqrt{\sqrt{1+\omega_L^2\tilde{\tau}_s^2}+1}}
{\sqrt{1+\omega_L^2\tilde{\tau}_s^2}}
=\left. \frac{\tilde{L}}{\left(2\pi\right)^{1/2}}f_r(\omega_L) \right|_{\tilde{L}\ll 1}.
\end{equation}
Naturally, it contains only a single scale $\omega_L\sim \tilde{\tau}_s^{-1}$.

If the magnetization  of the injector is along the $x$-axis while the magnetization of the detector is along the $y$-axis, then  the Hanle signal is proportional to $S_y(t)$.
The corresponding expression for nonlocal resistance reads
\begin{multline}
\label{tilde-three-terms}
\hspace*{-1cm}
\tilde{R}(B) \propto \Biggl\{\frac{B_1^2}{2 H^2} f_i\left(y_{\s \omega} \right)
 +  \frac{1}{2} \left( 1 - \frac{B_1^2}{2 H^2} + \frac{B_0-\frac{\omega}{\gamma}}{H} \right)f_i\left(y_{\s \omega + \gamma H} \right) \\
- \frac{1}{2} \left( 1 - \frac{B_1^2}{2 H^2} - \frac{B_0-\frac{\omega}{\gamma}}{H} \right)f_i\left(y_{\s - \omega +\gamma H}\right)\Biggr\},
\end{multline}
where the function $f_i(y)$ is defined through Eq. (\ref{propto}) as $\text{Im} f(y)$,
which amounts to the change of cosine by sine in the right-hand-side.
In the absence of drive, the resistance $\tilde{R}(B)$ is an odd function of magnetic field.   In the limit of small $\tilde{L}$ it simplifies to
\begin{equation}
\label{smallLtilde}
\tilde{R}(B)\propto \frac{\sqrt{\sqrt{1+\omega_L^2\tilde{\tau}_s^2}-1}}
{\sqrt{1+\omega_L^2\tilde{\tau}_s^2}}
=\left. \frac{\tilde{L}}{\left(2\pi\right)^{1/2}}f_i(\omega_L) \right|_{\tilde{L}\ll 1}.
\end{equation}

\begin{figure}
\includegraphics[width=90mm]{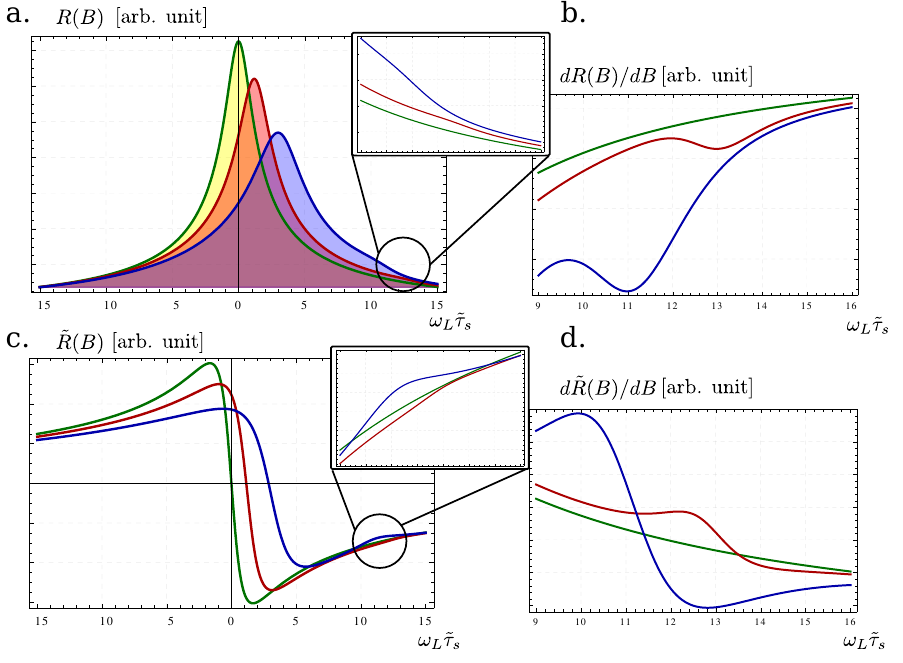}
\caption{[Color online] (a) Upon increasing the drive amplitude the Hanle curve broadens and its maximum is shifted quadratically with $B_1$. Green, red, and blue curves are plotted from Eq. (\ref{three-terms}) dimensionless drive frequency $\omega\tilde{\tau}_s=7$ and three dimesionless drive amplitudes: $\gamma B_1\tilde{\tau_s}=0$ (green curve) $\gamma B_1\tilde{\tau_s}=4$ (red curve), and $\gamma B_1\tilde{\tau_s}=6$ (blue curve). Magnetic resonance
$\omega_L=\omega$ resides around $\omega_L\tilde{\tau}_s=11$. The corresponding region is enlarged. Magnetic resonance is pronounced in the $dR/dB$ shown in (b); (c) and (d) are the same as (a) and (b) for the nonlocal resistance $\tilde{R}(B)$.  }
\label{lsmall}
\end{figure}

To find the shape of the Hanle curve in the presence of drive, the asymptote Eq. (\ref{smallL}) should be substituted into Eq. (\ref{three-terms}). In Fig. \ref{lsmall} we plot the modified Hanle curves calculated for the driving frequency $\omega\tilde{\tau}_s=7$ and two magnitudes of the drive
$\gamma B_1 \tilde{\tau}_s=4$ and $\gamma B_1 \tilde{\tau}_s=6$. We also plot the corresponding
curves for $\tilde{R}(B)$. The chosen value of
$\omega$ is so big because the FWHM value of  $R(B)$ in the absence of drive is also big, approximately $4/\tilde{\tau}_s$. One can identify in Fig. \ref{lsmall}
three major features caused by the drive:

\noindent{(i)} The shift of the maximum. The origin of this shift is the interplay of the prefactor and the function $f_r\left(y_{\s - \omega +\gamma H}\right)$ in the third term of Eq. (\ref{three-terms}). Firstly, this term gives the dominant contribution to $R(B)$ for small $B_1$. This is because the prefactor in the first term is $\propto B_1^2$, while the prefactor in the second term is $\propto B_1^4$ for $\omega_L < \omega$.    On the other hand, the prefactor in the third term changes rapidly from $1$ to $0$ at $\omega_L=\omega$.
With regard to $f_r\left(y_{\s - \omega +\gamma H}\right)$, it has two peaks at
\begin{equation}
\omega_L =\omega_{\pm}= \omega \pm \sqrt{\omega^2-\left(\gamma B_1\right)^2}.
\end{equation}
The  peak at $\omega_{+}$  is eliminated by the prefactor, while the peak at $\omega_{-}$, which behaves as
$(\gamma B_1)^2/2\omega$ at small $B_1$, survives and defines the  position of maximum in $R(B)$. For the two driving amplitudes plotted in Fig. \ref{lsmall} the expected shifts of the maxima
are related as $9:4$, which is indeed the case.

\noindent{(ii)} The Hanle curves broaden with increasing the drive amplitude. Formally, this
follows from the broadening of the step-like behavior of the prefactor in the third term with
$B_1$.

\noindent{(iii)} Upon increasing $B_1$, the Hanle curves exhibit signatures of magnetic resonance. True magnetic resonance, $\omega_L=\omega$, is certainly present only in  the dynamics of $S_z$.
In the dynamics of $S_x$ and $S_y$ the manifestations of magnetic resonance is vague and
originates from the fact that  the derivative of the arguments $\omega\pm \gamma H$ with
respect to $B$ is equal to $(\omega \pm \gamma B_0)/\gamma H$. It passes
through zero at the magnetic-resonance condition and changes rapidly  from $-1$ to $1$ in its vicinity. This change translates into a kink-like behavior indicated in Fig. \ref{lsmall}.
More pronounced signatures of the magnetic resonance can be seen in the derivative $dR/dB$ also
shown in Fig. \ref{lsmall}. The derivative develops a plateau.

The shapes of $\tilde{R}(B)$ shown in Fig. \ref{lsmall} evolve with the drive in a predictable
fashion. Namely, the position of zero shifts to finite magnetic field $\sim (\gamma B_1)^2/2\omega$ and the curves broaden. The signatures of magnetic resonance are more
pronounced in $\tilde{R}(B)$. As seen in Fig. \ref{lsmall} the derivative $d\tilde{R}/dB$ exhibits a jump-like behavior near $\omega_L=\omega$.

\subsection{The injector and the detector are far apart}
\begin{figure}
\includegraphics[width=90mm]{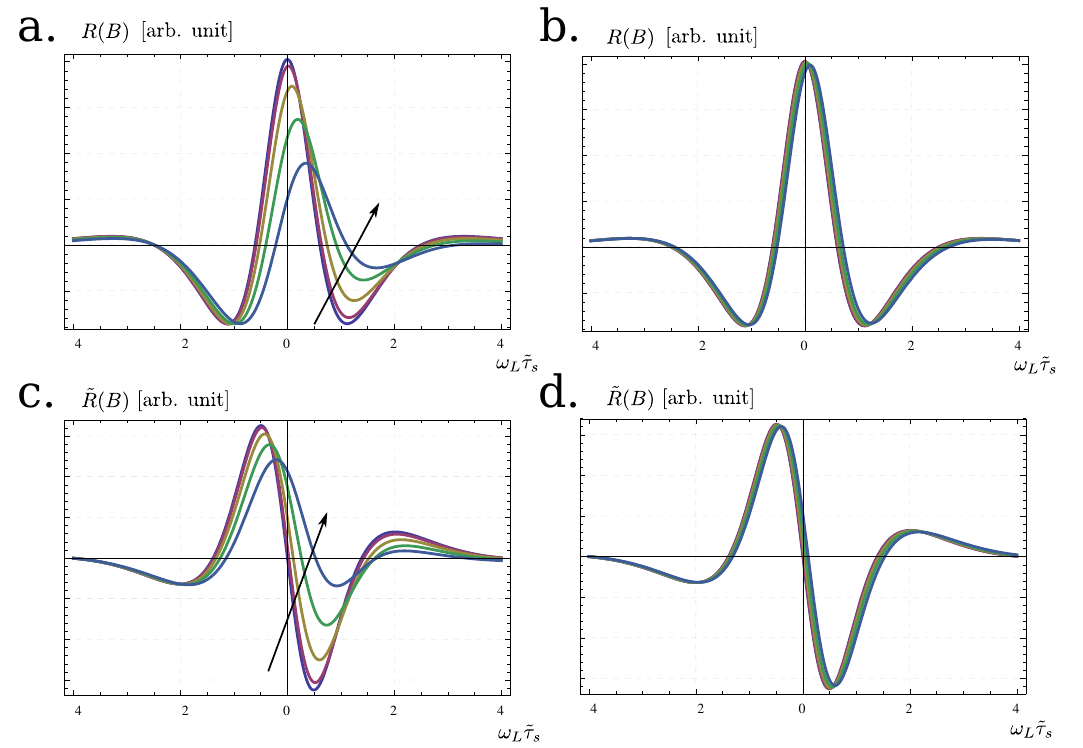}
\caption{[Color online] (a) In a spin-transport device with a long channel the prime effect of the ac drive on the Hanle curve is elimination of a minimum.
The $R(B)$ curves are plotted from Eq. (\ref{three-terms}) for $L/\lambda_s =2.4$, $\omega\tilde{\tau}_s =1.1$ and five values: $0$,
 $0.25$, $0.5$, $0.75$, and $1$ of the driving field, $\gamma B_1\tilde{\tau}_s$. Black arrow shows the direction in which $\gamma B_1$ increases. (b) Same curves as (a) but for the high-frequency drive,
 $\omega\tilde{\tau}_s=5$, show a weak response to the drive. (c) and (d) are the same as (a) and (b) for $\tilde{R}(B)$.   }
\label{lbig}
\end{figure}
In a long device, $\tilde{L}\gg 1$,  the nonlocal resistance Eq. (\ref{propto})
exhibits oscillations  decaying with magnetic field. First two zeros correspond to magnetic
fields $\omega_L\tilde{\tau}_s \approx \pi/2\tilde{L}$ and  $\omega_L\tilde{\tau}_s \approx 3\pi/2\tilde{L}$. The effect of the ac drive on $R(B)$ is most pronounced when the
driving frequency lies between these two values. This is illustrated in Fig. \ref{lbig}.
Two sets of curves in Fig. \ref{lbig} correspond to the same values of the drive amplitudes
but to different driving frequencies. In the left and right sets the frequency differ by a factor of $5$. It is seen that the Hanle shapes in the right set do not respond to the drive. The reason for
that is that the value $\omega\tilde{\tau}_s$ for this set is $5$, which is much bigger than
$\pi/\tilde{L}\approx 1.4$. For the left set,  $\omega\tilde{\tau}_s=1.1$, which is close to
$\pi/\tilde{L}$. The lively response of $R(B)$ and $\tilde{R}(B)$ to the drive at this frequency originates from the fact that a non-driven curve is flat around $\omega_L\tilde{\tau}_s \approx 1.2$. With the choice  $\omega\tilde{\tau}_s=1.1$ this $\omega_L$ is near magnetic resonance and fast change of the prefactors in Eq. (\ref{three-terms}) with $\omega_L$ is not overshadowed by the change of the function $f_r(y)$.

\section{Linear polarization of the drive}
\begin{figure}
\includegraphics[width=77mm]{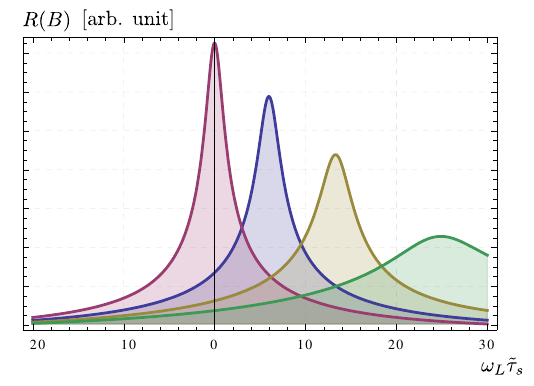}
\caption{[Color online] Evolution of the Hanle curves in a short device with increasing the amplitude of drive and very high driving frequency, $\omega\tilde{\tau}_s=30$, The curves are plotted from Eq.
(\ref{three-terms})
for the values of $\gamma B_1\tilde{\tau}_s$: $0$ (magenta), $18$ (blue), $25$ (tan), and $30$ (green). The prime effect of drive is the shift of the maximum with slow broadening of the shape.}
\label{movingright}
\end{figure}
The expressions for nonlocal resistance obtained in the previous section are exact,
in the sense, that they apply at arbitrary strengths and frequencies of the circularly polarized drive.
We analyzed them for the situation when both $\gamma B_1$ and $\omega$ are comparable
to the width of the Hanle curve. It is easy to see from Eq. (\ref{three-terms}) what happens to the Hanle curve when both  $\gamma B_1$ and $\omega$ are much bigger than $\gamma B_0$. With the shape of the Hanle curve dominated by the third term in
Eq. (\ref{three-terms}), the argument $-\omega +\gamma H$ of $f_r$ in this term
can, at low $B_0$, be   expanded as
\begin{equation}
\label{expanded}
-\omega+\gamma H \approx -\omega+\Bigl[(\gamma B_1)^2+\omega^2\Bigr]^{1/2} -
\frac{\gamma B_0 \omega}{\Bigl[(\gamma B_1)^2+\omega^2\Bigr]^{1/2}}.
\end{equation}
Eq. (\ref{expanded}) suggests that under a fast
and strong circularly-polarized drive the Hanle curve
simply shifts to the right preserving its shape. This is illustrated
in Fig. \ref{movingright} where the $R(B)$ curves are plotted from Eq. (\ref{three-terms}) for fast and strong drives. We intentionally chose a very high driving frequency   $\omega\tilde{\tau}_s =30$ to allow the $R(B)$ peak
to shift substantially with increasing  $\gamma B_1\tilde{\tau}_s$.

Obviously, under a linearly polarized drive, the $R(B)$~--~dependence maintains its symmetry with respect to $B~=~0$. From Fig. \ref{movingright} one would expect that,
when the maximum of $R(B)$ for circularly polarized drive is shifted by more
than the width in the absence of drive, then the effect of linearly polarized
drive would be a ``symmetrized"  peak. This is, actually, not the case. The reason is that the spin dynamics for a fast linearly polarized and circularly
polarized drives are very different.

Assume that the driving field oscillates along the $x$--axis, ${\bf B}_1(t)={\bf i}B_{1}\cos(\omega t+\varphi)$. Then  equations of
motion for the spin projections assume the form
\begin{align}
\frac{dS_x}{dt} &= - \gamma B_0 S_y, \\
\frac{dS_y}{dt} &=   \gamma B_0 S_x -  \gamma B_1 S_z \cos \omega t, \\
\frac{dS_z}{dt} &= \gamma B_1 S_y \cos \omega t.
\end{align}
To handle the fast linearly polarized drive it is convenient\cite{modulated},
to switch to the variables
\begin{align}
S_{x'} &= S_x, \\
S_{y'} &= S_y \cos \theta(t) + S_z \sin \theta(t) , \\
S_{z'} &= -S_y \sin \theta(t) + S_z \cos \theta(t),
\end{align}
where the angle $\theta(t)$ is defined as
\begin{equation}
\theta(t)= \frac{\gamma B_1}{\omega} \sin (\omega t + \varphi)
\end{equation}
The physical meaning of the above transformation is moving into the frame rotating around the ac field; the ac field is ``canceled" in the new frame.
The equations of motion for the new variables read
\begin{align}
\label{InNewVariables}
\frac{dS_{x'}}{dt} &= \gamma B_0 S_{z'} \sin \theta(t) - \gamma  B_0 S_{y'} \cos \theta(t) , \\
\frac{dS_{y'}}{dt} &= \gamma B_0 S_{x'} \cos \theta(t), \\
\frac{dS_{z'}}{dt} &= -\gamma B_0 S_{x'} \sin \theta(t).
\end{align}
One can see that there are two natural frequencies in the system Eq. (\ref{InNewVariables}), one is $\gamma B_0$ and the other is $\omega$.
Since the second frequency is much bigger than the first, we can average
the equations over time interval $\left(-\frac{\pi}{\omega}, \frac{\pi}{\omega}\right)$ assuming that the spin projections do not change
significantly during this interval. Taking into account that $\langle \cos(\theta)\rangle = J_0(\gamma B_1/\omega)$, where $J_0(z)$ is a zero-order Bessel function,  we get
\begin{align}
\label{averaged}
\frac{dS_{x'}}{dt} &=  - \gamma  B_0 J_0\left(\frac{\gamma B_1}{\omega}\right) S_{y'} , \\
\frac{dS_{y'}}{dt} &= \gamma B_0 J_0\left(\frac{\gamma B_1}{\omega}\right) S_{x'}, \\
\frac{dS_{z'}}{dt} &= 0.
\end{align}
We see that the dynamics after averaging is slow, which justifies the averaging performed\cite{modulated}.
Upon returning to the lab frame the solution of the system Eq. (\ref{averaged}) satisfying the condition $S_x(0)=1$ reads
\begin{align}
\label{labframe}
S_x(t) &=  \cos\left[ \gamma B_0 J_0 \left( \frac{\gamma B_1}{\omega} \right) t \right], \\
S_y(t) &= \sin\left[ \gamma B_0 J_0 \left( \frac{\gamma B_1}{\omega} \right) t \right]
\cos \left[  \frac{\gamma B_1}{\omega} \sin(\omega t + \varphi) \right], \\
S_z(t) &= \sin\left[ \gamma B_0 J_0 \left( \frac{\gamma B_1}{\omega} \right) t \right]
\sin \left[  \frac{\gamma B_1}{\omega} \sin(\omega t + \varphi) \right].
\end{align}
As a final step, we average over the initial phase, $\varphi$, and obtain
\begin{align}
\label{final}
\langle S_x(t) \rangle  &=  \cos\left[ \gamma B_0 J_0 \left( \frac{ \gamma B_1}{\omega} \right) t \right], \\
\langle S_y(t) \rangle  &= J_0 \left( \frac{\gamma B_1}{\omega} \right) \sin\left[ \gamma B_0 J_0 \left( \frac{\gamma B_1}{\omega} \right) t \right], \\
\langle S_z(t) \rangle &= 0.
\end{align}
The above result leads us to the conclusion  that, with fast linearly polarized drive, the curves $R(B)$ and $\tilde{R}(B)$ have exactly the same shape as in the absence of drive. The only difference is that the Larmour frequency, $\omega_L$ gets replaced by $\omega_LJ_0(\gamma B_1/\omega)$, signifying the
broadening of the curves, which {\em oscillates} with the drive amplitude.

\subsection{Strong  drive}
Another regime of the spin dynamics specific for a linearly polarized drive is realized when the drive is very strong, $B_1 \gg B_0$. We will describe this regime qualitatively.
As ${\bf B}_1(t)$ oscillates, it exceeds the static field $B_0$ during, practically, the entire period,
$2\pi/\omega$. Then $B_0$ has a negligible effect on the spin dynamics.
However, during short time intervals, $\delta t$, when ${\bf B}_1(t)$ passes through zero, the electron spin is affected by $B_0$ only. During each of these intervals the
spin rotates by the angle $\sim B_0\delta t$. Thus, the net rotation after time $t$ is $\sim \left(B_0\delta t\right)\omega t$. Now the value $\delta t$ can be estimated
from the relation $B_1(\omega \delta t)=B_0$. This leads us to the conclusion that  the spin dynamics, averaged over the the period of drive, is still a regular spin precession
around the $z$--axis but with effective frequency $\sim \gamma B_0^2/B_1$ instead of $\omega_L$. One  consequence of the replacement of $\omega_L$ by $\gamma B_0^2/B_1 \ll \omega_L$ in Eq. (\ref{propto}) is a general broadening of the Hanle profile, which can be controlled by the strength of the drive. The other consequence is that the Hanle profile acquires a flat top.

\section{Discussion}
\begin{figure}
\includegraphics[width=77mm]{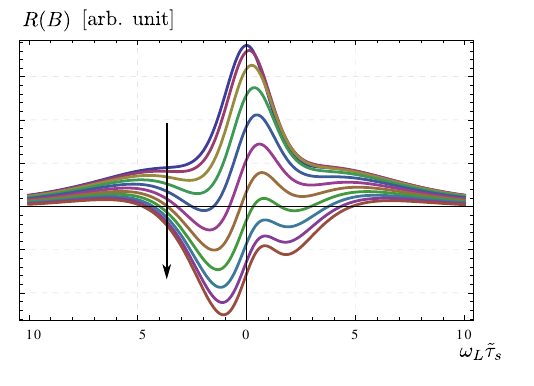}
\caption{At strong drive $\gamma B_1 > \omega$. The zero-drive Hanle curves not only
loose their symmetry but aquire additional maxima and minima. The evolution of the Hanle
shapes for a given $\gamma B_1\tilde{\tau_s}=2$ and $L/\lambda_s=1.4$ is plotted from
Eq. (\ref{three-terms}) for driving frequencies $\omega\tilde{\tau_s}$ taking the values between
$0$ and $1$ with a step $0.1$. Arrow shows the direction of the increase of the driving frequency.
}
\label{fancy}
\end{figure}
\begin{itemize}

\item

Our overall conclusion is that the ac drive with frequency, $\omega$,
affects  nonlocal spin transport if its amplitude is strong enough,
$\gamma B_1 \gtrsim \left(\frac{\omega}{\tilde{\tau}_s}\right)^{-1/2}$.
Choosing for an estimate the values, $\tau_s =50$~ns and $\omega=200$~MHz,
we find that the driving amplitude must be bigger than $1$~mT to affect
the spin-transport. This value is quite realistic for experiments where the effects of the ac drive are studied by electrical measurements\cite{BOEHME}.

\item

In this paper we considered the domain of parameters $\gamma B_1 < \omega$
and found that modification of the shapes of the Hanle curves is primarily
the broadening and the shift of the maximum. The above numerical estimate
suggests that the opposite relation, $\gamma B_1>\omega$ is also experimentally accessible. For this domain of a strong circularly polarized drive the shapes of  the Hanle curves change dramatically, as it is illustrated in Fig. \ref{fancy}. The curves exhibit {\em two} scales, which can be qualitatively interpreted as follows.
The low-$B$ scale is the signature of the condition $\omega_L=\omega$ for which the argument $y_{\s -\omega+\gamma H}$ of $f_r$ in Eq. (\ref{three-terms}) is minimal. The high-$B$ feature is  the signature of $\omega_L=\gamma B_1$ at which the prefactor in the third term in Eq. (\ref{three-terms}) changes significantly.

\item

One of our main findings is that the fast and strong circularly polarized ac field
affect the Hanle profile dramatically, as illustrated in Fig. \ref{movingright},
while the linearly polarized field with the same amplitude and frequency has a little effect on the Hanle curve. This is in contrast to usual reasoning in magnetic resonance suggesting to treat the linear polarization as superposition of two circular polarizations and keep only the polarization which co-rotates with precessing electron spin.  The easiest way to understand the difference is the spin
dynamics for linear and circular polarizations qualitatively is to set $B_0$ equal to zero. Then, with initial spin direction along $x$, the  ac field, linearly polarized along $x$, would not cause any spin dynamics {\em at all}. On the other hand, according to Eqs. (\ref{Sx}) and (\ref{Sy}), the average spin will precess around $z$--axis (if $\gamma B_1 \ll \omega$) even when $B_0=0$.

\item

 The effect of the ac drive on the Hanle curve is more pronounced for circular polarization of the drive when
 the symmetry of the curve is broken. In experiment,
 the circular polarization of microwaves is achieved\cite{CircularMicrowave,CircularForNV,CircularPolarization}
with the help of two crossed microstrip resonators.
In particular, in Ref. (\onlinecite{CircularForNV}) it was demonstrated
that ODMR spectrum of the nitrogen vacancies in diamond depends on the
direction of the circular polarization of  microwaves.

\item

Sensitivity of the Hanle curves to the ac-drive can serve as additional proof
that a  spin-polarized current  indeed flows through the channel of the device.
This proof is especially important for three-terminal devices where the question
about the spin injection is still controversial\cite{InvertedHanleSi, Jansen3T,Other3T}.
In these devices, with only one of electrodes being a ferromagnet, the spin-dependent buildup of
a voltage between injector and detector observed in experiment, and even sensitivity of this voltage to the magnetic field,\cite{InvertedHanleSi, Jansen3T,Other3T}
can be caused by different delicate mechanisms. For example, a small inhomogeneity of
the magnetization of the electrode might cause this sensitivity. Obviously, this source of
magnetoresistance would not be sensitive to the ac-drive.

%

\end{itemize}



\section{Acknowledgements}
  We are grateful to C. Boehme and H. Malissa for insightful discussions.
 This work was supported by NSF through MRSEC DMR-1121252.

\end{document}